\documentclass[pra,aps,twocolumn,superscriptaddress, longbibliography, notitlepage]{revtex4-1}
\usepackage[colorlinks=true, citecolor=blue, urlcolor=blue, linkcolor=blue]{hyperref}
\usepackage{graphicx}
\usepackage{dcolumn}

\usepackage{xcolor}
\usepackage{cancel}

\usepackage{bm}
\usepackage{amsmath,amsfonts}
\usepackage[english]{babel}
\usepackage{amsthm}
\usepackage{hyperref}
\usepackage{xcolor}
\usepackage{enumerate}
\usepackage{braket}
\usepackage{physics}
\theoremstyle{plain}

\newtheorem{definition}{Definition}
\newtheorem{lemma}[definition]{Lemma}
\newtheorem{proposition}[definition]{Proposition}
\newtheorem{corollary}[definition]{Corollary}
\newtheorem{theorem}[definition]{Theorem}

\begin{document}
\title{Measurement-device-independent entanglement witness with imprecise input states}
\author{Xing-Chen Guo}
\affiliation{School of Mathematics, South China University of Technology, GuangZhou 510640, China}
\author{Mao-Sheng Li}
\email{li.maosheng.math@gmail.com}
\affiliation{School  of Mathematics, South China University of Technology, GuangZhou 510640, China}

\begin{abstract}
Measurement-device-independent entanglement witnesses (MDI-EWs) enable the detection of entanglement without relying on characterized measurements. However, the entanglement criteria for MDI-EWs typically assume the idealized condition that the lab input states are precisely the desired ones. In this work, we remove this idealization by considering the realistic scenario in which lab input states may be imprecise. We introduce a new class of MDI-EWs, termed sensitive MDI-EWs, whose entanglement criteria fail under any non-zero imprecision in the input states. We derive sufficient conditions for an MDI-EW to be classified as sensitive, revealing that a large class of MDI-EWs exhibit this sensitivity. Additionally, we demonstrate that two well-known MDI-EWs for Werner states are sensitive according to our criteria, one of which recovers the result from a previous study [\href{https://doi.org/10.1103/PhysRevA.104.012429}{Phys. Rev. A \textbf{104}, 012429 (2021)}]. Moreover, we clarify the concept of lab input states in a way that makes imprecisions experimentally measurable, and propose a systematic approach for modifying the criterion of any MDI-EW to accommodate small imprecisions, thus enhancing experimental relevance. A simple MDI-EW example is provided, where our modified criterion is already optimal. This work   bridges the gap between idealized theoretical models and practical experimental conditions, paving the way for more robust and accessible entanglement detection in real-world settings.
\end{abstract}

\maketitle

\section{INTRODUCTION}
The significance of entanglement in quantum information science makes the detection of entanglement an important issue \cite{1,2,3}. Entanglement witnesses(EWs) as observables allow us to detect entanglement directly from the measurement statistics, and are widely used for entanglement detection in the lab \cite{2,3,4,5,6}. But to detect entanglement via EWs, we shall ensure that the lab measurements are precisely the desired ones, which is hard to realize. Indeed, even small measurement imprecision can invalidate the certification of entanglement by EWs \cite{7,8}. To deal with this, a strategy is proposed, where the notion of measurement imprecision is formalized as experimentally accessible quantity and by which the entanglement criteria from EWs can be modified to account for the imprecise measurements \cite{9}. This strategy also applies to steering detection \cite{10,11}. Experimental detection of genuine multipartite entanglement with imprecise measurements has also been shown recently \cite{12}.\par

Another way to detect entanglement is through the so-called Measurement-Device-Independent Entanglement Witnesses(MDI-EWs) obtained from the EWs \cite{13}. This method can certify entanglement from the measurement statistics regardless of the measurements performed \cite{13}. However, in contrast to EWs, the entanglement criteria of MDI-EWs demand that the lab input states are precisely the desired ones. Given that the lab input states are inevitably imprecise, it is a must to study the effects of the imprecise input states on the criteria of MDI-EWs, and modify the criteria in the presence of imprecise input states if necessary, as has been done to account for measurement imprecisions in the case of EWs \cite{13}. So far, the effects of some input states noises on the criteria of MDI-EWs are known, where the entanglement criterion of the famous MDI-EW for Werner states is shown to be invalid with some noisy input states \cite{14}. But a systematic approach to modify the criterion of any MDI-EW by experimentally accessible input states imprecisions is still lacking, which is known for the case of EWs. This work is devoted to bridge the gap.\par

In this work, we introduce the concept of sensitive MDI-EW, whose entanglement criterion becomes invalid in the presence of any non-zero input states imprecision. Practical criteria for an MDI-EW to be sensitive are given, from which we can see that a large class of MDI-EWs are sensitive. We then apply the criteria to show that two famous MDI-EWs for Werner states are sensitive \cite{13}, one of which recovers the result of  \cite{14}.  The sensitivity phenomenon shows the experimental need to modify the entanglement criteria of MDI-EWs to account for the imprecise input states, even if the lab input states are nearly the desired ones. To address this issue, we clarify the notion of lab input states, which allows for the quantification and operational estimation of input states imprecisions in the lab. Based on this, we propose a systematic approach to modify the entanglement criterion of any MDI-EW, the modified criterion can detect entanglement in a near-optimal manner when the imprecisions are small, which is relevant to the experiment. Finally we exemplify the approach by a simple MDI-EW, in which our modification is already optimal.\par

The remainder of the article is organized as follows. Section \ref{s2} reviews MDI-EWs and introduces the terminology used throughout the article. Section \ref{s3} presents the introduction and analysis of sensitive MDI-EWs. Section \ref{s4} discusses how to modify the criterion of any MDI-EW to account for imprecise lab input states. Finally, Section \ref{s5} concludes the article with remarks on future research directions.

\section{MEASUREMENT-DEVICE-INDEPENDENCE ENTANGLEMENT WITNESSES}\label{s2}
We begin by reviewing the definition of MDI-EWs and their basic properties, as well as introducing the conventions and notations used in this article.
\par

In this work, we consider weakly optimal entanglement witnesses, specifically Hermitian operators $\widetilde{W}$ acting on a finite-dimensional Hilbert space $H_A\otimes H_B$. These operators satisfy $\min\limits_{\sigma\in\Omega_s}\Tr(\widetilde{W}\sigma)=0$ and $\Tr(\widetilde{W}\rho)<0$ for some entangled state $\rho$, where $\Omega_s$ denotes the set of all separable states \cite{2}. It should be noted that general entanglement witnesses can be optimized to yield optimal witnesses 
 \cite{15}, which makes weakly optimal witnesses of particular interest. Therefore, we focus on this case.  We can decompose $\widetilde{W}$ as
\[\widetilde{W}=\sum\limits_{x,y}\beta_{x,y}\widetilde{\tau}_x^T\otimes\widetilde{\omega}_y^T,\]
where $\beta_{x,y}$ are real numbers, $\widetilde{\tau}_x$ and $\widetilde{\omega}_y$ are pure states  acting on $H_A$ and $H_B$, respectively. \par

Now consider the following experiment(semi-quantum nonlocal game\cite{16}): Alice and Bob share an unknown bipartite state $\rho$ acting on $H_A\otimes H_B$, where Alice's state acts on $H_A$ and Bob's acts on $H_B$. When Alice (Bob) inputs $x(y)$, she (he) attaches the state $\tau_x$ ($\omega_y$) acting on $H_A^{'}$ ($H_B^{'}$) to her (his) share of $\rho$, and then performs a measurement on the composite system using an uncharacterised device. The conditional probability $p_{\rho}(a,b\lvert x,y)$ is given by  
\[p_{\rho}(a,b\lvert x,y)=\Tr(A_a\otimes B_b\cdot\tau_x\otimes\rho\otimes\omega_y),\]
for some POVM operators $A_a$, $B_b$ acting on $H_A^{'}\otimes H_A$, $H_B\otimes H_B^{'}$ respectively. For any pure separable state $\sigma=\ket{\psi}\bra{\psi}\otimes\ket{\phi}\bra{\phi}$ acting on $H_A\otimes H_B$, we have
\begin{align*}
\sum\limits_{x,y}\beta_{x,y}p_{\sigma}(a,b\lvert x,y)&=\sum\limits_{x,y}\beta_{x,y}\Tr(\tau_x\otimes\omega_y\cdot A\otimes B)\\
&=\sum\limits_{x,y}\beta_{x,y}\langle A\rangle_{\tau_x}\langle B\rangle_{\omega_y},
\end{align*}
where $A=\Tr_{H_A}(A_a\cdot I\otimes\ket{\psi}\bra{\psi})$, $B=\Tr_{H_B}(B_b\cdot \ket{\phi}\bra{\phi}\otimes I)$ are POVM operators acting on $H_A^{'}$, $H_B^{'}$ respectively. Hereafter, we omit the underlying Hilbert spaces. Note that $A$ and $B$ can be arbitrary POVM operators, as long as  the POVM operators $A_a$, $B_b$ are chosen  arbitrarily (e.g., by selecting $A_a=A\otimes I$, $B_b=I\otimes B$ for any $A$, $B$).\par

Now suppose that $\tau_x=\widetilde{\tau}_x$, $\omega_y=\widetilde{\omega}_y$ for each $x, y$. Then we have
\begin{align*}
\min\sum\limits_{x,y}\beta_{x,y}p_{\sigma}(a,b\lvert x,y)&=\min\sum\limits_{x,y}\beta_{x,y}\Tr(\widetilde{\tau}_x\otimes\widetilde{\omega}_y\cdot A\otimes B)\\
&=\min\Tr(\widetilde{W}\cdot A^T\otimes B^T)\\
&=0,
\end{align*}
where the first minimization is over all separable states $\sigma$ and all POVM operators $A_a$, $B_b$; the second and the third minimizations are over all POVM operators $A$, $B$. Hence, the observation $$\sum\limits_{x,y}\beta_{x,y}p_{\rho}(a,b\lvert x,y)<0$$ indicates that $\rho$ is entangled regardless of the measurements performed. For a real matrix $(\beta_{x,y})\neq\mathbf{0}$ and two sets of states $\{\widetilde{\tau}_x\}$, and $\{\widetilde{\omega}_y\}$ (with the ranges of the indices 
$x$ and 
$y$ omitted for simplicity), we say $((\beta_{x,y}),\{\widetilde{\tau}_x\},\{\widetilde{\omega}_y\},0)$ is an MDI-EW in the sense that $\sum\limits_{x,y}\beta_{x,y}\widetilde{\tau}_x^T\otimes\widetilde{\omega}_y^T\equiv\widetilde{W}$ is entanglement witness.That is, in the corresponding experiment where the input states for $x, y$ are $\widetilde{\tau}_x$, $\widetilde{\omega}_y$ respectively, we can conclude that $\rho$ is entangled when $\sum\limits_{x,y}\beta_{x,y}p_{\rho}(a,b\lvert x,y)<0$.

\section{SENSITIVE MDI-EW}\label{s3}
Although the corresponding experiment of MDI-EW $((\beta_{x,y}),\{\widetilde{\tau}_x\},\{\widetilde{\omega}_y\},0)$ can certify entanglement when $\sum\limits_{x,y}\beta_{x,y}p_{\rho}(a,b\lvert x,y)<0$, it is an idealization to assume that the realistic experiment carried out in the laboratory matches precisely the one corresponding to $((\beta_{x,y}),\{\widetilde{\tau}_x\},\{\widetilde{\omega}_y\},0)$. Specifically, the sets of input states $\{\tau_x\}$ and $\{\omega_y\}$ in the realistic experiment may not be exactly the desired $\{\widetilde{\tau}_x\}$ and $\{\widetilde{\omega}_y\}$. In this scenario, we say the input states are \textit{imprecise}, and use the term \textit{imprecision} to refer to the difference between the realistic $\{\tau_x\}$, $\{\omega_y\}$ and the desired $\{\widetilde{\tau}_x\}$, $\{\widetilde{\omega}_y\}$.\par
It is natural to ask whether
$\sum\limits_{x,y}\beta_{x,y}p_{\rho}(a,b\lvert x,y)<0$ remains a valid criterion for certifying entanglement when the input states are imprecise, without misidentifying a separable state as entangled. The answer is no, as demonstrated by the seminal MDI-EW for Werner states in \cite{14}. But what about other MDI-EWs?\par
Indeed, we discover that for a large class of MDI-EWs, any non-zero imprecision would make the criterion $\sum\limits_{x,y}\beta_{x,y}p_{\rho}(a,b\lvert x,y)<0$ invalid to detect entanglement, no matter how small the imprecision is. To show this, we quantify the imprecision by the parameter $\epsilon$($0\leq\epsilon\leq1$) that constrains the sets of laboratory input states $\{\tau_x\}$, $\{\omega_y\}$ by the conditions $\Tr(\tau_x\widetilde{\tau}_x), \Tr(\omega_y\widetilde{\omega}_y)\geq1-\epsilon$ ($\forall x,y$). When $\epsilon=0$, the lab input states coincide exactly with the desired ones; while $\epsilon=1$ means that we only know the dimension of the lab input states. Now we introduce a class of MDI-EWs named sensitive MDI-EWs.

\begin{definition}[Sensitive MDI-EW]\label{definition1}
An MDI-EW $((\beta_{x,y}),\{\widetilde{\tau}_x\},\{\widetilde{\omega}_y\},0)$ is called sensitive if for any $0<\epsilon\leq1$, there exist sets of states $\{\tau_x\}$, $\{\omega_y\}$ with $\Tr(\tau_x\widetilde{\tau}_x), \Tr(\omega_y\widetilde{\omega}_y)\geq1-\epsilon$ for all $x$ and $y$, such that $((\beta_{x,y}),\{\tau_x\},\{\omega_y\},0)$ is no long an  MDI-EW. 
\end{definition}

Intuitively, \textit{sensitive} means that any non-zero imprecision, no matter how small it is, would invalidate the entanglement criterion of the MDI-EW. Mathematically, an MDI-EW $((\beta_{x,y}),\{\widetilde{\tau}_x\},\{\widetilde{\omega}_y\},0)$ is sensitive if and only if for $\forall\ 0<\epsilon\leq1$,  there exists POVM operators $A_a$, $B_b$ and $\{\tau_x\}$, $\{\omega_y\}$ (with $\Tr(\tau_x\widetilde{\tau}_x), \Tr(\omega_y\widetilde{\omega}_y)\geq1-\epsilon$, \  $\forall x,y$) and a separable state $\sigma$, such that 
\begin{align*}
\sum\limits_{x,y}\beta_{x,y}p_{\sigma}(a,b\lvert x,y)&=\sum\limits_{x,y}\beta_{x,y}\Tr(A_a\otimes B_b\cdot\tau_x\otimes\sigma\otimes\omega_y)\\
&<0.
\end{align*}
Equivalently,  $\forall\ 0<\epsilon\leq1$,  there exist POVM operators $A$, $B$  and a set of states $\{\tau_x\}$, $\{\omega_y\}$ (with $\Tr(\tau_x\widetilde{\tau}_x), \Tr(\omega_y\widetilde{\omega}_y)\geq1-\epsilon$, \ $\forall x,y$)  such that
\[
\sum\limits_{x,y}\beta_{x,y}\langle A\rangle_{\tau_x}\langle B\rangle_{\omega_y}<0.
\]

In addition, in the steering scenario, there are steering witnesses that remain valid for sufficiently small but non-zero measurement imprecision, exhibiting the phenomenon known as \textit{imprecision plateaus}\cite{11,17}. In the view of this, sensitive MDI-EWs are precisely the ones that do not support imprecision plateaus in the MDI-EW scenario.\par

Now we will show that a large class of MDI-EWs are sensitive. To do this, we need the following lemma, which might also be instructive to exhibit the ``sensitive" nature of some entanglement witnesses with imprecise measurements.    

\begin{lemma}\label{lemma2}
Let $\widetilde{E}$ be a POVM operator and $\widetilde{\tau}_1,\cdots,\widetilde{\tau}_m$ be pure states such that the inequality $\lambda_{\min}(\widetilde{E})<\langle{\widetilde{E}}\rangle_{\widetilde{\tau}_x} \leq \lambda_{\max}(\widetilde{E})<1$ holds for $x=1,\cdots,k$ where $k\leq m$. Then for any $ 0<\epsilon\leq1$ and any $\mathbf{e}=(e_1 ,\cdots ,e_m)^T\in\mathbb{R}^m$ with $e_{k+1},\cdots,e_{m}\geq0$, there exists some $y_0>0$, such that for any $0<y\leq y_0$,  the following relation holds:
\begin{equation}\label{eq:Difference}
 \left(
\begin{array}{c}
\langle E\rangle_{\tau_1}\\
\vdots \\
\langle E\rangle_{\tau_m}\\
\end{array}
\right)=\left(
\begin{array}{c}
\langle \widetilde{E}\rangle_{\widetilde{\tau}_1}\\
\vdots\\
\langle \widetilde{E}\rangle_{\widetilde{\tau}_m}\\
\end{array}
\right)+y\mathbf{e}
\end{equation}
 for some POVM operator $E$ and states $\tau_1,\cdots,\tau_m$ with $\Tr(\tau_x\widetilde{\tau}_x)\geq1-\epsilon$ for all $x=1,\cdots,m$.
\end{lemma}

Lemma \ref{lemma2} asserts that small perturbations of $(\langle \widetilde{E}\rangle_{\widetilde{\tau}_1},\cdots,\langle\widetilde{E}\rangle_{\widetilde{\tau}_m})^T$ in certain directions $\mathbf{e}$ can always be realized by $(\langle E\rangle_{\tau_1},\cdots,\langle E\rangle_{\tau_m})^T$, where $\{\tau_x\}$ can be chosen to be arbitrarily close to, but not equal to, $\{\widetilde{\tau}_x\}$. The statement of Lemma \ref{lemma2} is intuitively plausible, and the rigorous proof is presented in Appendix \ref{ap1}.\par

By definition, We know that for any MDI-EW $((\beta_{x,y}),\{\widetilde{\tau}_x\},\{\widetilde{\omega}_y\},0)$, there exist POVM operators $\widetilde{A}$ and $\widetilde{B}$ such that 
\begin{align*}
\sum\limits_{x,y}\beta_{x,y}\langle\widetilde{A}\rangle_{\widetilde{\tau}_x}\langle\widetilde{B}\rangle_{\widetilde{\omega}_y}&=\left(
         \begin{array}{c}
          \langle \widetilde{A}\rangle_{\widetilde{\tau}_1} \cdots \langle \widetilde{A}\rangle_{\widetilde{\tau}_m}\\
         \end{array}
          \right)(\beta_{x,y})
         \left(
          \begin{array}{c}
        \langle \widetilde{B}\rangle_{\widetilde{\omega}_1}\\
         \vdots\\
         \langle \widetilde{B}\rangle_{\widetilde{\omega}_n}\\
        \end{array}
        \right)\\
        &=0,
\end{align*}
where, without loss of generality, we assume that $(\beta_{x,y})\in\mathbb{R}^{m\times n}$, $\{\widetilde{\tau}_x\}=\{\widetilde{\tau}_x\}_{x=1}^{m}$, and $\{\widetilde{\omega}_y\}=\{\widetilde{\omega}_y\}_{y=1}^{n}$. To demonstrate that $((\beta_{x,y}),\{\widetilde{\tau}_x\},\{\widetilde{\omega}_y\},0)$ is sensitive, we need to find POVM operators $A$, $B$ and sets of states $\{\tau_x\}$, $\{\omega_y\}$, under arbitrary small but non-zero imprecision, such that
\begin{align*}
\sum\limits_{x,y}\beta_{x,y}\langle A\rangle_{\tau_x}\langle B\rangle_{\omega_y}&=\left(
         \begin{array}{c}
          \langle A\rangle_{\tau_1} \cdots \langle A\rangle_{\tau_m}\\
         \end{array}
          \right)(\beta_{x,y})
         \left(
          \begin{array}{c}
        \langle B\rangle_{\omega_1}\\
         \vdots\\
         \langle B\rangle_{\omega_n}\\
        \end{array}
        \right)\\
        &<0.
\end{align*}
To achieve this, it is natural to perturb $(\langle \widetilde{A}\rangle_{\widetilde{\tau}_1}\cdots\langle\widetilde{A}\rangle_{\widetilde{\tau}_m})^T$ and $(\langle \widetilde{B}\rangle_{\widetilde{\omega}_1}\cdots\langle\widetilde{B}\rangle_{\widetilde{\omega}_n})^T$ in suitable directions by replacing $\widetilde{A}$, $\widetilde{B}$ with $A$, $B$ and $\{\widetilde{\tau}_x\}$, $\{\widetilde{\omega}_y\}$ with $\{\tau_x\}$, $\{\omega_y\}$ under arbitrary small but non-zero imprecision.  Lemma \ref{lemma2} that tells us to what extent this can be done. This leads us to the following proposition, whose proof is presented in Appendix \ref{ap2}.

\begin{proposition}\label{p3}
An MDI-EW $((\beta_{x,y}),\{\widetilde{\tau}_x\},\{\widetilde{\omega}_y\},0)$ is sensitive, if there exist POVM operators $\widetilde{A}$, $\widetilde{B}$ that satisfy one of the following two conditions:
\begin{enumerate}[\rm(i)]
		         \item $\Tr(\widetilde{\tau}_x\otimes\widetilde{\omega}_y\cdot \widetilde{A}\otimes\widetilde{B})=0$ for each $x, y$;
              \item $\sum\limits_{x,y}\beta_{x,y}\Tr(\widetilde{\tau}_x\otimes\widetilde{\omega}_y\cdot\widetilde{A}\otimes\widetilde{B})=0$ and there exist $i, j$ such that $\beta_{i,j}\Tr(\widetilde{\tau}_i\otimes\widetilde{\omega}_j\cdot\widetilde{A}\otimes\widetilde{B})\neq0$.
\end{enumerate}
\end{proposition}

Proposition \ref{p3} provides a sufficient condition for an MDI-EW to be sensitive, revealing a large class of sensitive MDI-EWs. Note that from the condition (\rm{ii}) of Proposition \ref{p3}, for an MDI-EW $((\beta_{x,y}),\{\widetilde{\tau}_x\},\{\widetilde{\omega}_y\},0)$, if we find a separable state $\sigma$ such that 
\[
\sum\limits_{x,y}\beta_{x,y}\Tr(\widetilde{\tau}_x^T\otimes\widetilde{\omega}_y^T\cdot\sigma)=\Tr(\widetilde{W}\sigma)=0
\]
and $\beta_{i,j}\Tr(\widetilde{\tau}_i^T\otimes\widetilde{\omega}_j^T\cdot\sigma)\neq0$ for some $i, j$, then MDI-EW $((\beta_{x,y}),\{\widetilde{\tau}_x\},\{\widetilde{\omega}_y\},0)$ must be sensitive. This makes Proposition \ref{p3} a practical criterion for certifying sensitivity when we know the exact form of the separable state $\sigma$ that reaches $\Tr(\widetilde{W}\sigma)=0$.
\par

In particular, for an MDI-EW $((\beta_{x,y}),\{\widetilde{\tau}_x\},\{\widetilde{\omega}_y\},0)$ with $\beta_{x,y}\neq0$ for all $x$ and $y$, there always exist POVM operators $\widetilde{A}$, $\widetilde{B}$ such that $$
\sum\limits_{x,y}\beta_{x,y}\Tr(\widetilde{\tau}_x\otimes\widetilde{\omega}_y\cdot\widetilde{A}\otimes\widetilde{B})=0$$
by the definition of an MDI-EW. Thus, either condition $(\rm{i})$ or $(\rm{ii})$ of Proposition \ref{p3} holds, implying that the MDI-EW must be sensitive. This leads to the following practical criterion for determining the sensitivity of an MDI-EW:
\begin{corollary}\label{c4}
An MDI-EW $((\beta_{x,y}),\{\widetilde{\tau}_x\},\{\widetilde{\omega}_y\},0)$ is sensitive if $\beta_{x,y}\neq0$ for each x and y.
\end{corollary}

Now we apply Proposition \ref{p3} and Corollary \ref{c4} to demonstrate the sensitivity of two celebrated MDI-EWs for Werner states \cite{13,18,19,20}. Consider the weakly optimal entanglement witness for Werner states:
\[
\widetilde{W}=\frac{1}{2}\mathbb{I}-\ket{\Psi^-}\bra{\Psi^-},
\]
where $\ket{\Psi^-}=\frac{\ket{01}-\ket{10}}{\sqrt{2}}$, and $\Tr(\widetilde{W}\rho_{\frac{1}{3}})=0$ for separable Werner state $\rho_{\frac{1}{3}}=\frac{1}{3}\ket{\Psi^-}\bra{\Psi^-}+\frac{2}{3}\cdot\frac{\mathbb{I}}{4}$. To construct MDI-EWs, consider the following two different decompositions of $\widetilde{W}$, each leading to an MDI-EW. The first one is
\[
\widetilde{W}=\sum\limits_{x=0}^{3}\sum\limits_{y=0}^{3}\beta_{x,y}\widetilde{\tau}_x^T\otimes\widetilde{\omega}_y^T,
\]
where $\beta_{x,y}=\frac{5}{8}$ if $x=y$ and $\beta_{x,y}=-\frac{1}{8}$ otherwise, and $\widetilde{\tau}_x=\sigma_x\frac{\mathbb{I}+\vec{n}\cdot\vec{\sigma}}{2}\sigma_x$, $\widetilde{\omega}_y=\sigma_y\frac{\mathbb{I}+\vec{n}\cdot\vec{\sigma}}{2}\sigma_y$ with $\vec{n}=\frac{1}{\sqrt{3}}(1,1,1)$, $\vec{\sigma}=(\sigma_1,\sigma_2,\sigma_3)$ the vector of Pauli matrices, and $\sigma_0=\mathbb{I}$ the identity matrix. By Corollary \ref{c4}, it is evident that the corresponding MDI-EW $((\beta_{x,y}),\{\widetilde{\tau}_x\},\{\widetilde{\omega}_y\},0)$ is sensitive for $\beta_{x,y}\neq0$ ($\forall\ x,y$). Note that this result is consistent with Ref.\cite{14}, where the construction implicitly demonstrates the sensitivity of $((\beta_{x,y}),\{\widetilde{\tau}_x\},\{\widetilde{\omega}_y\},0)$. Next consider the second decomposition
\[
\widetilde{W}=\sum\limits_{x,y}\beta_{x,y}^{'}\widetilde{\tau}_x^{'T}\otimes\widetilde{\omega}_y^{'T},
\]
where $x, y$ are defined by $x=(x_1,x_2)$, $y=(y_1,y_2)$ with $x_1,y_1\in\{0,1\}$ and $x_2,y_2\in\{1,2,3\}$; and $\beta_{x,y}^{'}=\delta_{x_2,y_2}\frac{3\delta_{x_1,y_1}-1}{6}$ with Kronecker delta $\delta_{i,j}$; and $\widetilde{\tau}_x^{'}=\frac{\mathbb{I}+(-1)^{x_1}\sigma_{x_2}}{2}$, $\widetilde{\omega}_y^{'}=\frac{\mathbb{I}+(-1)^{y_1}\sigma_{y_2}}{2}$. Note that
\begin{align*}
&\beta_{(0,3),(0,3)}\Tr(\widetilde{\tau}_{(0,3)}^{'T}\otimes\widetilde{\omega}_{(0,3)}^{'T}\cdot\rho_{\frac{1}{3}})\\
=&\frac{1}{3}\Tr(\ket{0}\bra{0}\otimes\ket{0}\bra{0}\cdot\rho_{\frac{1}{3}})\\
=&\frac{1}{18}\neq 0, 
\end{align*}
together with the fact that $\Tr(\widetilde{W}\rho_{\frac{1}{3}})=0$, we conclude that the MDI-EW $((\beta_{x,y}^{'}),\{\widetilde{\tau}_x^{'}\},\{\widetilde{\omega}_y^{'}\},0)$ is sensitive by the condition $(\rm{ii})$ of Proposition \ref{p3}.

\section{MODIFIED MDI-EW WITH IMPRECISE INPUT STATES}\label{s4}
We now know that many MDI-EWs, including some well-known ones, are sensitive. Specifically, the criterion $$\sum\limits_{x,y}\beta_{x,y}p_{\rho}(a,b\lvert x,y)<0$$  fails to certify entanglement in the presence of any nonzero imprecision.  Given that we are typically dealing with small but non-zero imprecisions, it is crucial to find a systematic approach to modify MDI-EWs so that imprecise input states can be accounted for, even when these lab input states are nearly the desired ones.\par

To this end, we first model the lab input states as mixtures of the desired input states and some unknown states. For an MDI-EW $((\beta_{x,y}),\{\widetilde{\tau}_x\},\{\widetilde{\omega}_y\},0)$, the lab input states take the following form:
\begin{equation}\label{m1}
\begin{array}{rll}
&\tau_x=(1-p_A)\widetilde{\tau}_x+p_A\rho_{x}^{A},\\[2mm]
&\omega_y=(1-p_B)\widetilde{\omega}_y+p_B\rho_{y}^{B}
\end{array}
\end{equation}
for any $x$ and $y$, where $0\leq p_A,p_B\leq1$ and $\rho_{x}^{A}$, $\rho_{y}^{B}$ are some unknown states. When $p_A=p_B=0$, the lab input states are precisely the desired ones, while when $p_A=p_B=1$, the lab input states are completely unknown. From the constructions in the proof of Proposition \ref{p3}, the criterion in Eq.\eqref{m1} can be invalidated if $p_A, p_B>0$, highlighting the need for a modification of the criterion. Notably, in Ref. \cite{14}, the effects of various noisy input states on the MDI-EW for Werner states are explored, and these noisy states can be considered as special cases of 
Eq. \eqref{m1}  for some specific $\rho_x^{A}$, $\rho_y^{B}$.\par

Now we claim that this characterization of lab input states is relevant to the experiments, and that $p_A$ and $p_B$ can be estimated in the lab. The detection of entanglement requires access to the probabilities $\{p_{\rho}(a,b\lvert x,y)\ \lvert a,b,x,y\}$, which demands inputting each $x$ and $y$ multiple times, consuming a number of lab input states for each $x, y$. While it is realistic that in some runs of the experiment the lab input states are not the desired ones, we can still bound that probabilities with confidence by repeating the preparations of the input states and testing. Assume that the probability $p(\tau_x\neq\widetilde{\tau}_x)\leq p_A$ for any $x$, and similarly $p(\omega_y\neq\widetilde{\omega}_y)\leq p_B$ for any $y$, then
\begin{align*}
&p_{\rho}(a,b\lvert x,y)\\[2mm]
=&\sum\limits_{\tau_x,\omega_y}p_{\rho}(a,b\lvert\tau_x,\omega_y)\cdot p(\tau_x)\cdot p(\omega_y)\\
=&\Tr(A_a\otimes B_b\cdot\sum\limits_{\tau_x}p(\tau_x)\tau_x\otimes\rho\otimes\sum\limits_{\omega_y}p(\omega_y)\omega_y)\\
=&p_{\rho}(a,b\lvert (1-p_A)\widetilde{\tau}_x+p_A\rho_x^{A},(1-p_B)\widetilde{\omega}_y+p_B\rho_y^{B})
\end{align*}
holds for some states $\rho_x^{A}$, $\rho_y^{B}$. That is, we can equivalently take the lab input states in each run of the experiment as \eqref{m1} for each $x, y$. Though the exact forms of $\rho_x^{A}$  and $\rho_y^{B}$ are hard to determine, the values of $p_A$ and $p_B$ can still be estimated experimentally, allowing us to modify the MDI-EWs accordingly.\par

To modify MDI-EWs with respect to Eq. \eqref{m1} , we start with the extreme case where $p_A=p_B=1$, meaning the lab input states are completely unknown. In this case, we define 
\[
\chi((r_{x,y}))\equiv\min\limits_{\substack{a_x,b_y\in\{0,1\}}}\sum\limits_{x,y}r_{x,y}a_x b_y
\]
for any real matrix $(r_{x,y})$. Note that $\chi((r_{x,y}))$ can be easily computed when the matrix $(r_{x,y})$ is given. Hereafter we abbreviate $\chi((\beta_{x,y}))$ as $\chi$ when the underlying matrix $(\beta_{x,y})$ is clear. Next we introduce the following proposition, whose proof is given in Appendix \ref{ap3}.   

\begin{proposition}\label{p5}
Let $(\beta_{x,y})$ be a real matrix. Then
\[\min\sum\limits_{x,y}\beta_{x,y}\Tr(A_a\otimes B_b\cdot\tau_x\otimes\sigma\otimes\omega_y)=\chi,\]
where the minimization is over all POVM operators $A_a$, $B_b$, all sets of states $\{\tau_x\}$, $\{\omega_y\}$, and all separable states $\sigma$. 
\end{proposition}

Proposition \ref{p5} indicates that $\chi$ is exactly $$\min\sum\limits_{x,y}\beta_{x,y}p_{\sigma}(a,b\lvert x,y),$$ where the minimization is over all correlations $\{p_{\sigma}(a,b\lvert x,y)\lvert\ a,b,x,y\}$ obtained by separable states in the semi-quantum nonlocal games. In particular, for any MDI-EW $((\beta_{x,y}),\{\widetilde{\tau}_x\},\{\widetilde{\omega}_y\},0)$, when the lab input states are completely unknown ($p_A=p_B=1$ in \eqref{m1} ), the lowest probable value of $\sum\limits_{x,y}\beta_{x,y}p_{\sigma}(a,b\lvert x,y)$ obtained by separable states should be $\chi$, which is strictly less than the original value $0$ because there exist $i$, $j$ such that $\beta_{i,j}<0$. This suggests that we can certify entanglement in a device-independent manner if $\sum\limits_{x,y}\beta_{x,y}p_{\rho}(a,b\lvert x,y)<\chi$ in the semi-quantum nonlocal game, but this phenomenon exists if and only if 
\begin{equation}\label{eq:LessChi}
\inf\limits_{\rho}\sum\limits_{x,y}\beta_{x,y}p_{\rho}(a,b\lvert x,y)<\chi,
\end{equation}
which is not true for general $(\beta_{x,y})$. Whether there exists a real matrix $(\beta_{x,y})$ that satisfies the inequality \eqref{eq:LessChi} is an open question. Nevertheless, the characteristic of $\chi$ still has implication for modifying the criteria of MDI-EWs when $p_A$, $p_B$ are nearly but not 0, which should be a primary focus for the relevance to the realistic experiment.\par

Now consider the general case where, for an MDI-EW $((\beta_{x,y}),\{\widetilde{\tau}_x\},\{\widetilde{\omega}_y\},0)$, the lab input states take the forms of Eq. \eqref{m1} with arbitrary $p_A$ and $p_B$ values between 0 and 1. In this case, we have the following theorem, the proof of which is shown in Appendix \ref{ap4}.

\begin{theorem}\label{t6}
Let $((\beta_{x,y}),\{\widetilde{\tau}_x\},\{\widetilde{\omega}_y\},0)$ be an MDI-EW, with the dimension of the underlying Hilbert space of $\{\widetilde{\tau}_x\} (\{\widetilde{\omega}_y\})$ being  $d_A (d_B)$. Consider the experiment where the lab input states for each $x, y$ are given by $\tau_x=(1-p_A)\widetilde{\tau}_x+p_A\rho_{x}^{A}$ and $\omega_y=(1-p_B)\widetilde{\omega}_y+p_B\rho_{y}^{B}$, where $\rho_x^{A}$, $\rho_y^{B}$ are arbitrary states and $0\leq p_A,p_B\leq1$. Then
\begin{equation}\label{m2}
\sum\limits_{x,y}\beta_{x,y}p_{\sigma}(a,b\lvert x,y)\geq(p_A+p_B-p_Ap_B)\chi
\end{equation}
holds for any correlation $\{p_{\sigma}(a,b\lvert x,y)\lvert\ a,b,x,y\}$ obtained by separable states in this case.  Moreover,  provided the condition $$(p_A+p_B-p_Ap_B)<\frac{1}{d_Ad_B\chi}\lambda_{\min}(\sum\limits_{x,y}\beta_{x,y}\widetilde{\tau}_x^T\otimes\widetilde{\omega}_y^T),$$  there exists an entangled state that violates inequality \eqref{m2}.  
\end{theorem}

The last statement indicates that the modified criterion
\begin{equation}\label{m3}
\sum\limits_{x,y}\beta_{x,y}p_{\rho}(a,b\lvert x,y)<(p_A+p_B-p_Ap_B)\chi
\end{equation}
can successfully detect some entanglement when $p_A$, $p_B$ are small enough. Specifically,  Theorem \ref{t6} provides a systematic way to modify the criteria of MDI-EWs when the lab input states are nearly, but not exactly, the desired ones. Furthermore, several features indicate that this is an appropriate approach, as outlined below.\par

First, the method of Theorem \ref{t6} is analytical and suitable for any MDI-EW, while $(p_A+p_B-p_Ap_B)\chi$ is simple to compute directly from experimentally accessible quantities. This makes the modified criterion \eqref{m3} an experimentally friendly entanglement detection criterion.\par

Moreover, the bound $(p_A+p_B-p_Ap_B)\chi$ in \eqref{m2}  is continuous in $p_A$ and $p_B$. When $p_A=p_B=0$, it recovers the original optimal bound $0$. Therefore, the bound $(p_A+p_B-p_Ap_B)\chi$   is nearly the optimal bound when $p_A$, $p_B$ are nearly 0, which is highly relevant to experimental conditions. Additionally, when $p_A=p_B=1$, the bound  again recovers the optimal bound $\min\sum\limits_{x,y}\beta_{x,y}p_{\sigma}(a,b\lvert x,y)=\chi$, making it consistent across different scenarios.\par

Last, to illustrate the relevance of this approach, we present a simple example.  We will show that in this example, the bound $(p_A+p_B-p_Ap_B)\chi$ in \eqref{m2} is exactly the optimal bound. Consider the entanglement witness
\[
\widetilde{W}=\mathbb{I}-\sigma_1\otimes\sigma_1-\sigma_3\otimes\sigma_3,
\]
where $\sigma_1$, $\sigma_3$ are Pauli matrices with conventional subscripts. The  $\widetilde{W}$ has the following decomposition
\begin{align*}
\widetilde{W}&=2\ket{01}\bra{01} +2\ket{10}\bra{10}+\ket{+-}\bra{+-}\\
&\ \ +\ket{-+}\bra{-+} -\ket{++}\bra{++} -\ket{--}\bra{--},
\end{align*}
 where $\ket{+}=\frac{\ket{0}+\ket{1}}{\sqrt{2}}$, $\ket{-}=\frac{\ket{0}-\ket{1}}{\sqrt{2}}$. Let the input states be defined as follows: $\widetilde{\tau}_1=\widetilde{\omega}_1=\ket{0}\bra{0}$, $\widetilde{\tau}_2=\widetilde{\omega}_2=\ket{1}\bra{1}$, $\widetilde{\tau}_3=\widetilde{\omega}_3=\ket{+}\bra{+}$, $\widetilde{\tau}_4=\widetilde{\omega}_4=\ket{-}\bra{-}$. Define the matrix $(\beta_{x,y})$ as
 \[(\beta_{x,y})=
    \begin{pmatrix}
    \ 0 &\ 2 & 0 & 0\\
    \ 2 &\ 0 & 0 & 0\\
    \ 0 &\ 0 & -1 & 1\\
    \ 0 &\ 0 & 1 & -1
    \end{pmatrix}
,\]
 where $1\leq x,y\leq4$. Then $\widetilde{W}=\sum\limits_{x,y}\beta_{x,y}\widetilde{\tau}_x^T\otimes\widetilde{\omega}_y^T$, $((\beta_{x,y}),\{\widetilde{\tau}_x\},\{\widetilde{\omega}_y\},0)$ is an MDI-EW. As we consider lab input states of forms \eqref{m1}, we can compute $\chi$ as $$\chi\equiv\min\limits_{\substack{a_x,b_y\in\{0,1\}}}\sum\limits_{x,y}\beta_{x,y}a_x b_y=-1.$$  By Theorem \ref{t6},
 \[
 \sum\limits_{x,y}\beta_{x,y}p_{\sigma}(a,b\lvert x,y)\geq-(p_A+p_B-p_Ap_B),
 \]
 which holds for any correlation $\{p_{\sigma}(a,b\lvert x,y)\lvert\ a,b,x,y\}$ obtained from separable states $\sigma$.
 
 However, set POVM operators $A=B=\ket{-}\bra{-}$, and choose the input states:
\begin{align*}
&\tau_1=(1-p_A)\ket{0}\bra{0}+p_A\ket{+}\bra{+},\\
&\omega_1=(1-p_B)\ket{0}\bra{0}+p_B\ket{+}\bra{+},\\
&\tau_2=(1-p_A)\ket{1}\bra{1}+p_A\ket{+}\bra{+},\\
&\omega_2=(1-p_B)\ket{1}\bra{1}+p_B\ket{+}\bra{+},\\
&\tau_3=\omega_3=\ket{+}\bra{+},\ 
 \tau_4=\omega_4=\ket{-}\bra{-}.
\end{align*}
In this case, we have
\[
\sum\limits_{x,y}\beta_{x,y}\langle A\rangle_{\tau_x}\langle B\rangle_{\omega_y}=-(p_A+p_B-p_Ap_B),
\]
which means there exist POVM operators $A_a$, $B_b$ and a separable state $\sigma'$ such that
\begin{align*}
\sum\limits_{x,y}\beta_{x,y}p_{\sigma'}(a,b\lvert x,y)&=\sum\limits_{x,y}\beta_{x,y}\Tr(A_a\otimes B_b\cdot\tau_x\otimes\sigma'\otimes\omega_y)\\
&=\sum\limits_{x,y}\beta_{x,y}\langle A\rangle_{\tau_x}\langle B\rangle_{\omega_y}\\
&=-(p_A+p_B-p_Ap_B).
\end{align*}
Consequently, 
\[
\min\sum\limits_{x,y}\beta_{x,y}p_{\sigma}(a,b\lvert x,y)=-(p_A+p_B-p_Ap_B),
\]
where the minimization is over all correlations $\{p_{\sigma}(a,b\lvert x,y)\lvert\ a,b,x,y\}$ obtained from separable states. This justifies that in this simple example, the modified bound $(p_A+p_B-p_Ap_B)\chi$ in Eq. \eqref{m2} is optimal, and therein the modified entanglement criterion \eqref{m3} is already the best possible for this case. 

\section{DISCUSSION} \label{s5}
We have characterized sensitive MDI-EWs and developed a systematic approach to modify the criterion of any MDI-EW to account for imprecise input states. This study indicates several promising directions for further investigation. One natural direction is to apply our modified MDI-EW criteria in experiments to detect entanglement under imprecise input states, particularly for the simple MDI-EW presented at the end of Sec. \ref{s4}. This could enable more practical and accurate experimental entanglement detection, even in the presence of imperfections in the input states. Another intriguing problem is whether there exists an MDI-EW that is not sensitive, i.e., one that supports an ``imprecision plateau". Given that our sufficient conditions for sensitivity are relatively weak, it is possible that all MDI-EWs are sensitive. This remains an open question that warrants further exploration. Building on Proposition \ref{p5}, an interesting question arises: does there exist a real matrix  $(\beta_{x,y})$ such that $$\inf\limits_{\rho}\sum\limits_{x,y}\beta_{x,y}p_{\rho}(a,b\lvert x,y)<\chi?$$ Considering that entanglement can be detected device-independently via Bell inequalities in nonlocal games \cite{21,22,23}, our question may provide insight for detecting entanglement device-independently in the more general semi-quantum nonlocal games \cite{16}. This could significantly extend our understanding of entanglement detection in scenarios where the detection devices themselves are untrusted. Finally, a natural next step is to consider the generalization of our results to multipartite systems. Expanding our findings to multi-party scenarios could provide new insights into multipartite entanglement detection and extend the applicability of MDI-EWs to more complex quantum systems.

\vskip 10pt

\begin{center}
    {\noindent{\bf ACKNOWLEDGMENTS}\, \,}
\end{center}

This work was supported by National Natural Science Foundation of China under Grants No. 12371458, the Guangdong Basic and Applied Basic Research Foundation under Grants Nos. 2023A1515012074, 2024A1515010380  and the Science and Technology Planning Project of Guangzhou under Grants No. 2023A04J1296.

\bigskip

\onecolumngrid

\appendix

\section{Proof of Lemma 2}\label{ap1}
\begin{proof}
The statement holds for the case $\mathbf{e}=\mathbf{0}$. Now we assume that $\mathbf{e}\neq\mathbf{0}$. Consider the spectral decomposition
\[\widetilde{E}=\lambda_1P_1+\cdots+\lambda_lP_l\]
where $\lambda_{\max}(\widetilde{E})=\lambda_1>\cdots>\lambda_l=\lambda_{\min}(\widetilde{E})$ are the eigenvalues of $\widetilde{E}$ and $P_1,\cdots,P_l$ are the corresponding projective operators. Define
\begin{equation}\label{eq:def_delta}\delta=\min\{1-\lambda_1, \min\limits_{x\atop \langle P_1\rangle_{\widetilde{\tau}_x}\neq0}\{\frac{\epsilon}{2}(\langle\widetilde{E}\rangle_{\widetilde{\tau}_x}-\lambda_l)\}\}.\end{equation}
Then $\delta>0$ for $\lambda_1=\lambda_{\max}(\widetilde{E})<1$ and when $\langle P_1\rangle_{\widetilde{\tau}_x}\neq0$ we have $\langle\widetilde{E}\rangle_{\widetilde{\tau}_x}>\lambda_l$. Next we define
\[E=\widetilde{E}+\delta P_1=(\lambda_1+\delta)P_1+\cdots+\lambda_lP_l.\]
Then $\mathbf{0}<E\leq\mathbf{I}$ as $\lambda_1+\delta\leq 1.$ Hence $E$ is a POVM operator. Now we choose states $\rho_1,\cdots,\rho_m$ such that
\begin{equation}\label{eq:threesettings}
\begin{cases}
\langle E\rangle_{\rho_x}=\lambda_l=\lambda_{\min}(\widetilde{E}), & \  \text{for those } x \text{ with  }   \langle P_1\rangle_{\widetilde{\tau}_x}\neq 0,\\
\langle E\rangle_{\rho_x}-\langle\widetilde{E}\rangle_{\widetilde{\tau}_x}=0, & \  \text{for those } x \text{ with  }   \langle P_1\rangle_{\widetilde{\tau}_x}=0  \text{ 
 and } e_x=0,\\
\frac{\langle E\rangle_{\rho_x}-\langle\widetilde{E}\rangle_{\widetilde{\tau}_x}}{e_x}>0, &  \  \text{for those } x \text{ with  }  \langle P_1\rangle_{\widetilde{\tau}_x}=0  \text{ 
 and }  e_x\neq0 .
\end{cases}
\end{equation}
Note that this can be done: since  $\lambda_l\leq \langle\widetilde{E}\rangle_{\widetilde{\tau}_x}\leq \lambda_1$ and $\langle E\rangle_{\rho_x}$ can be can be chosen arbitrarily between $\lambda_l$ and $\lambda_1+\delta$, then $\langle E\rangle_{\rho_x}-\langle\widetilde{E}\rangle_{\widetilde{\tau}_x}$ can be chosen to be strictly greater than 0 or equal to 0. If $e_x<0$, then by the condition $e_{k+1},...,e_m\geq0$, we have $x\in\{1,\cdots,k\}$ and thus $\lambda_l<\langle\widetilde{E}\rangle_{\widetilde{\tau}_x}\leq\lambda_1$, we can suitably choose $\rho_x$ such that $\langle E\rangle_{\rho_x}<\langle\widetilde{E}\rangle_{\widetilde{\tau}_x}$.\\
Fix $\rho_1,\cdots, \rho_m$ as above, define
\begin{equation}\label{eq:def_y0}y_0=\min\{\min\limits_{\substack{x\\ e_x\neq0 \\ \langle P_1\rangle_{\widetilde{\tau}_x}=0}}\{\frac{\epsilon(\langle E\rangle_{\rho_x}-\langle\widetilde{E}\rangle_{\widetilde{\tau}_x})}{e_x}\},\min\limits_{\substack{x\\e_x\neq0\\ \langle P_1\rangle_{\widetilde{\tau}_x}\neq0}}\{\frac{\delta\langle P_1\rangle_{\widetilde{\tau}_x}}{\lvert e_x\rvert}\}\}.
\end{equation}
Then $y_0>0$. For any $0<y\leq y_0$, define
\begin{equation}\label{eq:twosettings}
\begin{cases}
p_x=\frac{ye_x-\delta\langle P_1\rangle_{\widetilde{\tau}_x}}{\langle E\rangle_{\rho_x}-\langle E\rangle_{\widetilde{\tau}_x}}, & \text{ for those } x \text{ with }  \langle E\rangle_{\rho_x}-\langle E\rangle_{\widetilde{\tau}_x}\neq0, \\
p_x=\epsilon, & \text{ for those } x \text{ with }  \langle E\rangle_{\rho_x}-\langle E\rangle_{\widetilde{\tau}_x}=0.
\end{cases}    
\end{equation}
Now we claim that $0\leq p_x\leq\epsilon (x=1,\cdots,m)$. When $\langle E\rangle_{\rho_x}-\langle E\rangle_{\widetilde{\tau}_x}=0$, we have $0<p_x=\epsilon$. When $\langle E\rangle_{\rho_x}-\langle E\rangle_{\widetilde{\tau}_x}\neq0$: if $\langle P_1\rangle_{\widetilde{\tau}_x}=0$ which corresponds to the third case of Eq. \eqref{eq:threesettings}, we have
\[0\leq p_x=\frac{ye_x}{\langle E\rangle_{\rho_x}-\langle E\rangle_{\widetilde{\tau}_x}}=\frac{ye_x}{\langle E\rangle_{\rho_x}-\langle\widetilde{E}\rangle_{\widetilde{\tau}_x}}\leq\frac{y_0e_x}{\langle E\rangle_{\rho_x}-\langle\widetilde{E}\rangle_{\widetilde{\tau}_x}}\leq\epsilon;\]
if $\langle P_1\rangle_{\widetilde{\tau}_x}\neq0$ which corresponds to the first case of Eq. \eqref{eq:threesettings} and $\langle E\rangle_{\widetilde{\tau}_x}=\langle \widetilde{E}\rangle_{\widetilde{\tau}_x}+\delta \langle P_1\rangle_{\widetilde{\tau}_x}>\lambda_l$, we have
\[p_x=\frac{\delta\langle P_1\rangle_{\widetilde{\tau}_x}-ye_x}{\langle E\rangle_{\widetilde{\tau}_x}-\lambda_l}\geq\frac{\delta\langle P_1\rangle_{\widetilde{\tau}_x}-y_0\lvert e_x\rvert}{\langle E\rangle_{\widetilde{\tau}_x}-\lambda_l}\geq0\]
and
\[p_x=\frac{\delta\langle P_1\rangle_{\widetilde{\tau}_x}-ye_x}{\langle E\rangle_{\widetilde{\tau}_x}-\lambda_l}\leq\frac{\delta\langle P_1\rangle_{\widetilde{\tau}_x}+y_0\lvert e_x\rvert}{\langle E\rangle_{\widetilde{\tau}_x}-\lambda_l}\leq\frac{2\delta\langle P_1\rangle_{\widetilde{\tau}_x}}{\langle E\rangle_{\widetilde{\tau}_x}-\lambda_l}\leq\epsilon\]
by definitions of $y_0$ and $\delta$ respectively.
So $0\leq p_x\leq\epsilon(x=1,\cdots,m)$, $i.e.$, $1-\epsilon\leq1-p_x\leq1(x=1,\cdots,m)$. Now we define
\[\tau_x=(1-p_x)\widetilde{\tau}_x+p_x\rho_x, \quad x=1,\cdots,m.\]
Then clearly $\Tr(\tau_x\widetilde{\tau}_x)\geq1-p_x\geq1-\epsilon(x=1,\cdots,m)$. Finally, for any $x\in\{1,\cdots,m\}$, we have
\begin{align*}
\langle E\rangle_{\tau_x}-\langle\widetilde{E}\rangle_{\widetilde{\tau}_x}&=\langle E\rangle_{\tau_x}-\langle E\rangle_{\widetilde{\tau}_x}+\langle E\rangle_{\widetilde{\tau}_x}-\langle\widetilde{E}\rangle_{\widetilde{\tau}_x} \\&=\langle E\rangle_{\tau_x-\widetilde{\tau}_x}+\langle E-\widetilde{E}\rangle_{\widetilde{\tau}_x}\\
&=p_x(\langle E\rangle_{\rho_x}-\langle E\rangle_{\widetilde{\tau}_x})+\delta\langle P_1\rangle_{\widetilde{\tau}_x}\\
&=ye_x.
\end{align*}
To see the last equality holds, it suffices to consider the case $\langle E\rangle_{\rho_x}-\langle E\rangle_{\widetilde{\tau}_x}=0$. Note that in this case, we must have $\langle P_1\rangle_{\widetilde{\tau}_x}=0$, otherwise we would obtain $\langle E\rangle_{\rho_x}-\langle E\rangle_{\widetilde{\tau}_x}=\lambda_l-\langle\widetilde{E}\rangle_{\widetilde{\tau}_x}-\delta\langle P_1\rangle_{\widetilde{\tau}_x}<0$, contradiction. Adopting the fact that $\langle P_1\rangle_{\widetilde{\tau}_x}=0$, we get $\langle E\rangle_{\rho_x}-\langle \widetilde{E}\rangle_{\widetilde{\tau}_x}=\langle E\rangle_{\rho_x}-\langle E\rangle_{\widetilde{\tau}_x}=0$. This could happen only in the   second case of Eq. \eqref{eq:threesettings}. Hence  $e_x=0$. So we get $p_x(\langle E\rangle_{\rho_x}-\langle E\rangle_{\widetilde{\tau}_x})+\delta\langle P_1\rangle_{\widetilde{\tau}_x}=0=ye_x$.
\end{proof}

\section{Proof of Proposition 3}\label{ap2}
\begin{proof}
Assume that $x\in\{1,\cdots,m\}$ and $y\in\{1,\cdots,n\}$, that is, $(\beta_{x,y})\in\mathbb{R}^{m\times n}$, $\{\widetilde{\tau}_x\}=\{\widetilde{\tau}_x\}_{x=1}^{m}$, $\{\widetilde{\omega}_y\}=\{\widetilde{\omega}_y\}_{y=1}^{n}$. Now we prove the proposition by cases.
\begin{enumerate}[(a)]
\item $\Tr(\widetilde{\tau}_x\otimes\widetilde{\omega}_y\cdot \widetilde{A}\otimes\widetilde{B})=0$ for each $x, y$.
\paragraph*{} We have either $(\langle \widetilde{A}\rangle_{\widetilde{\tau}_1},\cdots,\langle \widetilde{A}\rangle_{\widetilde{\tau}_m})=\mathbf{0}$ or $(\langle\widetilde{B}\rangle_{\widetilde{\omega}_1},\cdots,\langle\widetilde{B}\rangle_{\widetilde{\omega}_n})=\mathbf{0}$, without loss of generality we assume that $(\langle \widetilde{A}\rangle_{\widetilde{\tau}_1}\cdots\langle \widetilde{A}\rangle_{\widetilde{\tau}_m})=\mathbf{0}$. First note that if $\sum\limits_{y=1}^{n}\beta_{x,y}\widetilde{\omega}_y$ is positive for any $x\in\{1,\cdots,m\}$, then we will have $\sum\limits_{x=1}^{m}\widetilde{\tau}_x\otimes\sum\limits_{y=1}^{n}\beta_{x,y}\widetilde{\omega}_y$ is positive semidefinite. But by the definition of MDI-EW, $$\widetilde{W}=\sum\limits_{x=1}^{m}\sum\limits_{y=1}^{n}\beta_{x,y}\widetilde{\tau}_x^T\otimes\widetilde{\omega}_y^T=(\sum\limits_{x=1}^{m}\widetilde{\tau}_x\otimes\sum\limits_{y=1}^{n}\beta_{x,y}\widetilde{\omega}_y)^T$$ is entanglement witness,  there  should exist an entangled state $\rho_e$ such that $\Tr\widetilde{W}\rho_e<0$. So $\widetilde{W}$  is not positive semidefinite.  This also implies that $\sum\limits_{x=1}^{m}\widetilde{\tau}_x\otimes\sum\limits_{y=1}^{n}\beta_{x,y}\widetilde{\omega}_y=\widetilde{W}^T $ is  not positive semidefinite.  Thus we would obtain a contradiction.  Therefore,  there must exist some $s\in\{1,\cdots,m\}$ such that $\sum\limits_{y=1}^{n}\beta_{s,y}\widetilde{\omega}_y$ is non-positive. Without loss of generality, we  assume that $\sum\limits_{y=1}^{n}\beta_{1,y}\widetilde{\omega}_y$ is non-positive. Then there exists a POVM operator $B$ such that $$\langle B\rangle_{\sum\limits_{y=1}^{n}\beta_{1,y}\widetilde{\omega}_y}=\sum\limits_{y=1}^{n}\beta_{1,y}\langle B\rangle_{\widetilde{\omega}_y}<0.$$  Now for any $0<\epsilon\leq1$, we choose $A=\widetilde{A}$, $\tau_1=(1-\epsilon)\widetilde{\tau}_1+\epsilon\frac{\widetilde{A}}{\Tr\widetilde{A}}$, $\tau_x=\widetilde{\tau}_x(x=2,\cdots,m)$, $\omega_y=\widetilde{\omega}_y(y=1,\cdots,n)$. Then $A$, $B$ are POVM operators and $\{\tau_x\}$, $\{\omega_y\}$ are sets of states that sastisfy $\Tr(\tau_x\widetilde{\tau}_x),\Tr(\omega_y\widetilde{\omega}_y)\geq1-\epsilon$ for any $x, y$. And
\[
\left(
\begin{array}{c}
\langle A\rangle_{\tau_1}, \cdots, \langle A\rangle_{\tau_m}\\
\end{array}
\right)(\beta_{x,y})
\left(
\begin{array}{c}
\langle B\rangle_{\omega_1}\\
\vdots\\
\langle B\rangle_{\omega_n}\\
\end{array}
\right)=\langle\widetilde{A}\rangle_{\tau_1}\sum\limits_{y=1}^{n}\beta_{1,y}\langle B\rangle_{\widetilde{\omega}_y}=\epsilon\frac{\Tr\widetilde{A}^2}{\Tr\widetilde{A}}\sum\limits_{y=1}^{n}\beta_{1,y}\langle B\rangle_{\widetilde{\omega}_y}<0,
\]
which means MDI-EW $((\beta_{x,y}),\{\widetilde{\tau}_x\},\{\widetilde{\omega}_y\},0)$ is sensitive.
\item $\sum\limits_{x,y}\beta_{x,y}\Tr(\widetilde{\tau}_x\otimes\widetilde{\omega}_y\cdot\widetilde{A}\otimes\widetilde{B})=0$ and there exists $i, j$ such that $\beta_{i,j}\Tr(\widetilde{\tau}_i\otimes\widetilde{\omega}_j\cdot \widetilde{A}\otimes\widetilde{B})\neq0$.

We separate the argument into  several cases according to $\lambda_{\min}(\widetilde{A})$ and $\lambda_{\min}(\widetilde{B})$.

   \begin{enumerate}[(i)]
        \item $\lambda_{\min}(\widetilde{A})=\lambda_{\min}(\widetilde{B})=0$. Since there exists $i, j$ such that $\beta_{i,j}\langle\widetilde{A}\rangle_{\widetilde{\tau}_i}\langle\widetilde{B}\rangle_{\widetilde{\omega}_j}\neq0$, we have $(\langle \widetilde{A}\rangle_{\widetilde{\tau}_1},\cdots,\langle \widetilde{A}\rangle_{\widetilde{\tau}_m})\neq\mathbf{0}$ and $(\langle\widetilde{B}\rangle_{\widetilde{\omega}_1},\cdots,\langle\widetilde{B}\rangle_{\widetilde{\omega}_n})\neq\mathbf{0}$. After relabelling the index $i,j$, then term $\sum\limits_{x=1}^{m}\sum\limits_{y=1}^{n}\beta_{x,y}\langle\widetilde{A}\rangle_{\widetilde{\tau}_x}\langle\widetilde{B}\rangle_{\widetilde{\omega}_y}$ can be written as 
        \begin{align*}
\sum\limits_{x=1}^{m}\sum\limits_{y=1}^{n}\beta_{x,y}\langle\widetilde{A}\rangle_{\widetilde{\tau}_x}\langle\widetilde{B}\rangle_{\widetilde{\omega}_y}
        &=\left(
         \begin{array}{c}
          \langle \widetilde{A}\rangle_{\widetilde{\tau}_1} ,\cdots ,\langle \widetilde{A}\rangle_{\widetilde{\tau}_m}\\
         \end{array}
          \right)(\beta_{x,y})
         \left(
          \begin{array}{c}
        \langle \widetilde{B}\rangle_{\widetilde{\omega}_1}\\
         \vdots\\
         \langle \widetilde{B}\rangle_{\widetilde{\omega}_n}\\
        \end{array}
        \right)
       \\ &=\left(
           \begin{array}{c}
           \widetilde{\alpha}^T \ \mathbf{0}
           \end{array}
           \right)
           \left(
           \begin{array}{c}
           M_{11}\ M_{12}\\
           M_{21}\ M_{22}
           \end{array}
           \right)
           \left(
           \begin{array}{c}
           \widetilde{\beta}\\ 
           \mathbf{0}
           \end{array}
           \right)\\
           &=\widetilde{\alpha}^T M_{11}\widetilde{\beta}\\
           &=0.
\end{align*}
Here the components of $\alpha$ are precisely the non-zero components of $(\langle\widetilde{A}\rangle_{\widetilde{\tau}_1},\cdots,\langle\widetilde{A}\rangle_{\widetilde{\tau}_m})$, similarly the components of $\widetilde{\beta}$ are the non-zero components of $(\langle\widetilde{B}\rangle_{\widetilde{\omega}_1},\cdots,\langle\widetilde{B}\rangle_{\widetilde{\omega}_n})$. Without loss of generality, we assume that $\widetilde{\alpha}=(\langle\widetilde{A}\rangle_{\widetilde{\tau}_1},\cdots,\langle\widetilde{A}\rangle_{\widetilde{\tau}_k})^T$ and $\widetilde{\beta}=(\langle\widetilde{B}\rangle_{\widetilde{\omega}_1},\cdots,\langle\widetilde{B}\rangle_{\widetilde{\omega}_l})^T$, where $k$, $l\geq1$. By the condition ``there exists $i, j$ such that $\beta_{i,j}\Tr(\widetilde{\tau}_i\otimes\widetilde{\omega}_j\cdot\widetilde{A}\otimes\widetilde{B})\neq0$", we must have $M_{11}\neq\mathbf{0}$. Now we claim that there exist an $\mathbf{e_\alpha}\in\mathbb{R}^k$ and  an $\mathbf{e_\beta}\in\mathbb{R}^l$, such that for any $y>0$ we have
\[
(\widetilde{\alpha}+y\mathbf{e_\alpha})^T M_{11}(\widetilde{\beta}+y\mathbf{e_\beta})=y\mathbf{e_\alpha}^T M_{11}\widetilde{\beta}+y\widetilde{\alpha}^T M_{11}\mathbf{e_\beta}+y^2\mathbf{e_\alpha}^T M_{11}\mathbf{e_\beta}<0.
\]
In fact,  note that if $\widetilde{\alpha}^T M_{11}\neq\mathbf{0}$, then we can always find an $\mathbf{e_\beta}\in\mathbb{R}^l$ such that $\widetilde{\alpha}^T M_{11}\mathbf{e_\beta}<0$.  For this case, set $\mathbf{e_\alpha}=\mathbf{0}$, then we obtain $(\widetilde{\alpha}+y\mathbf{e_\alpha})^T M_{11}(\widetilde{\beta}+y\mathbf{e_\beta})=y\widetilde{\alpha}^T M_{11}\mathbf{e_\beta}<0$. If $\widetilde{\alpha}^T M_{11}=\mathbf{0}$ and $M_{11}\widetilde{\beta}\neq\mathbf{0}$, there is always some  $\mathbf{e_\alpha}\in\mathbb{R}^k$ such that $\mathbf{e_\alpha}^T M_{11}\widetilde{\beta}<0$. For this case, set $\mathbf{e_\beta}=\mathbf{0},$ then we get $(\widetilde{\alpha}+y\mathbf{e_\alpha})^T M_{11}(\widetilde{\beta}+y\mathbf{e_\beta})=y\mathbf{e_\alpha}^T M_{11}\widetilde{\beta}<0$. Finally, 
if $\widetilde{\alpha}^T M_{11}=\mathbf{0}$ and $M_{11}\widetilde{\beta}=\mathbf{0}$, since $M_{11}\neq\mathbf{0}$, we can always find $\mathbf{e_\alpha}\in\mathbb{R}^k$ and $\mathbf{e_\beta}\in\mathbb{R}^l$ such that $\mathbf{e_\alpha}^T M_{11}\mathbf{e_\beta}<0$. So again $(\widetilde{\alpha}+y\mathbf{e_\alpha})^T M_{11}(\widetilde{\beta}+y\mathbf{e_\beta})=y^2\mathbf{e_\alpha}^T M_{11}\mathbf{e_\beta}<0$.

Now applying Lemma \ref{lemma2}, we can find $y_0>0$ and POVM operators $A$, $B$ and sets of states $\{\tau_x\}$, $\{\omega_y\}$ with $\Tr(\tau_x\widetilde{\tau}_x)$, $\Tr(\omega_y\widetilde{\omega}_y)\geq1-\epsilon$ for any $ x, y$ such that
\[
\left(
\begin{array}{c}
\langle A\rangle_{\tau_1}\\
\vdots \\
\langle A\rangle_{\tau_m}\\
\end{array}
\right)=\left(
\begin{array}{c}
\widetilde{\alpha}+y_0\mathbf{e_\alpha}\\
\\
\mathbf{0}\\
\end{array}
\right), \quad
\left(
\begin{array}{c}
\langle B\rangle_{\omega_1}\\
\vdots \\
\langle B\rangle_{\omega_n}\\
\end{array}
\right)=\left(
\begin{array}{c}
\widetilde{\beta}+y_0\mathbf{e_\beta}\\
\\
\mathbf{0}\\
\end{array}
\right).
\]
Hence 
\begin{align*}
        \sum\limits_{x=1}^{m}\sum\limits_{y=1}^{n}\beta_{x,y}\langle A\rangle_{\tau_x}\langle B\rangle_{\omega_y}
        &=\left(
         \begin{array}{c}
          \langle A\rangle_{\tau_1}, \cdots ,\langle A\rangle_{\tau_m}\\
         \end{array}
          \right)(\beta_{x,y})
         \left(
          \begin{array}{c}
        \langle B\rangle_{\omega_1}\\
         \vdots\\
         \langle B\rangle_{\omega_n}\\
        \end{array}
        \right)\\
        &=(\widetilde{\alpha}+y_0\mathbf{e_\alpha})^T M_{11}(\widetilde{\beta}+y_0\mathbf{e_\beta})\\
        &<0.
\end{align*}
So MDI-EW $((\beta_{x,y}),\{\widetilde{\tau}_x\},\{\widetilde{\omega}_y\},0)$ is sensitive.

\item $\lambda_{\min}(\widetilde{A})>0$ or $\lambda_{\min}(\widetilde{B})>0$. Without loss of generality, we assume that $\lambda_{\min}(\widetilde{A})>0$. Now we show that in this case, we have
\[
 \sum\limits_{x=1}^{m}\sum\limits_{y=1}^{n}\beta_{x,y}\langle A^{'}\rangle_{\widetilde{\tau}_x}\langle \widetilde{B}\rangle_{\widetilde{\omega}_y}
        =\left(
         \begin{array}{c}
          \langle A^{'}\rangle_{\widetilde{\tau}_1} ,\cdots ,\langle A^{'}\rangle_{\widetilde{\tau}_m}\\
         \end{array}
          \right)(\beta_{x,y})
         \left(
          \begin{array}{c}
        \langle \widetilde{B}\rangle_{\widetilde{\omega}_1}\\
         \vdots\\
         \langle \widetilde{B}\rangle_{\widetilde{\omega}_n}\\
        \end{array}
        \right)=0
\]
for any POVM operator $A^{'}$. Since $\lambda_{\min}(\widetilde{A})>0$, for any POVM operator $A^{'}$, there exists $\delta>0$ such that $\widetilde{A}-\delta A^{'}$ is POVM operator, then we have
\begin{align*}
\left(
         \begin{array}{c}
          \langle \widetilde{A}-\delta A^{'}\rangle_{\widetilde{\tau}_1} ,\cdots, \langle \widetilde{A}-\delta A^{'}\rangle_{\widetilde{\tau}_m}\\
         \end{array}
          \right)(\beta_{x,y})
         \left(
          \begin{array}{c}
        \langle \widetilde{B}\rangle_{\widetilde{\omega}_1}\\
         \vdots\\
         \langle \widetilde{B}\rangle_{\widetilde{\omega}_n}\\
        \end{array}
        \right)\geq0.
\end{align*}
for $((\beta_{x,y}),\{\widetilde{\tau}_x\},\{\widetilde{\omega}_y\},0)$ is MDI-EW. Subsequently,
\[
-\delta\left(
         \begin{array}{c}
          \langle A^{'}\rangle_{\widetilde{\tau}_1} ,\cdots, \langle A^{'}\rangle_{\widetilde{\tau}_m}\\
         \end{array}
          \right)(\beta_{x,y})
         \left(
          \begin{array}{c}
        \langle \widetilde{B}\rangle_{\widetilde{\omega}_1}\\
         \vdots\\
         \langle \widetilde{B}\rangle_{\widetilde{\omega}_n}\\
        \end{array}
        \right)\geq0.
\]
But we also have $\left(
         \begin{array}{c}
          \langle A^{'}\rangle_{\widetilde{\tau}_1} ,\cdots ,\langle A^{'}\rangle_{\widetilde{\tau}_m}\\
         \end{array}
          \right)(\beta_{x,y})
         \left(
          \begin{array}{c}
        \langle \widetilde{B}\rangle_{\widetilde{\omega}_1}\\
         \vdots\\
         \langle \widetilde{B}\rangle_{\widetilde{\omega}_n}\\
        \end{array}
        \right)\geq0$ for $A^{'}$ is POVM operator.  Hence,
\[
\left(
         \begin{array}{c}
          \langle A^{'}\rangle_{\widetilde{\tau}_1} ,\cdots ,\langle A^{'}\rangle_{\widetilde{\tau}_m}\\
         \end{array}
          \right)(\beta_{x,y})
         \left(
          \begin{array}{c}
        \langle \widetilde{B}\rangle_{\widetilde{\omega}_1}\\
         \vdots\\
         \langle \widetilde{B}\rangle_{\widetilde{\omega}_n}\\
        \end{array}
        \right)=0.
\]
Fixed  $\widetilde{\tau}_1,\cdots,\widetilde{\tau}_m$, there always some POVM $A'$ such that $0<\lambda_{\min}(A')<\langle A'\rangle_{\widetilde{\tau}_x}\leq\lambda_{\max}(A')<1(x\in\{1,\cdots,m\})$. By replacing $\widetilde{A}$ with $A'$,  we can always assume that $\widetilde{A}$ satisfies $\lambda_{\min}(\widetilde{A})<\langle\widetilde{A}\rangle_{\widetilde{\tau}_x}\leq\lambda_{\max}(\widetilde{A})<1(x\in\{1,\cdots,m\})$.

Also, if $\lambda_{\min}(\widetilde{B})>0$, we can without loss of generality assume that $\lambda_{\min}(\widetilde{B})<\langle\widetilde{B}\rangle_{\widetilde{\omega}_y}\leq\lambda_{\max}(\widetilde{B})<1$ holds for any $y\in\{1,\cdots,n\})$. Since $(\beta_{x,y})\neq\mathbf{0}$, then similar to the argument in the case (b)(i), there exist $\mathbf{e_\alpha}\in\mathbb{R}^m$ and $\mathbf{e_\beta}\in\mathbb{R}^n$, such that for any $y>0$ we have
\[
(\widetilde{\alpha}+y\mathbf{e_\alpha})^T (\beta_{x,y})(\widetilde{\beta}+y\mathbf{e_\beta})=y\mathbf{e_\alpha}^T (\beta_{x,y})\widetilde{\beta}+y\widetilde{\alpha}^T (\beta_{x,y})\mathbf{e_\beta}+y^2\mathbf{e_\alpha}^T (\beta_{x,y})\mathbf{e_\beta}<0,
\]
here $\widetilde{\alpha}=(\langle \widetilde{A}\rangle_{\widetilde{\tau}_1}, \cdots, \langle \widetilde{A}\rangle_{\widetilde{\tau}_m})^T$ and $\widetilde{\beta}=(\langle \widetilde{B}\rangle_{\widetilde{\omega}_1} ,\cdots ,\langle \widetilde{B}\rangle_{\widetilde{\omega}_n})^T$. Then following the routine of case (b)(i), we can deduce that MDI-EW $((\beta_{x,y}),\{\widetilde{\tau}_x\},\{\widetilde{\omega}_y\},0)$ is sensitive by applying Lemma \ref{lemma2}.

If $\lambda_{\min}(\widetilde{B})=0$, then  without loss of generality  we assume that $\langle\widetilde{B}\rangle_{\widetilde{\omega}_1},\cdots,\langle\widetilde{B}\rangle_{\widetilde{\omega}_l}>0$ and $\langle\widetilde{B}\rangle_{\widetilde{\omega}_{l+1}},\cdots,\langle\widetilde{B}\rangle_{\widetilde{\omega}_n}=0$, where $l\geq1$. Write
 \begin{align*}
        \sum\limits_{x=1}^{m}\sum\limits_{y=1}^{n}\beta_{x,y}\langle\widetilde{A}\rangle_{\widetilde{\tau}_x}\langle\widetilde{B}\rangle_{\widetilde{\omega}_y}
        &=\left(
         \begin{array}{c}
          \langle \widetilde{A}\rangle_{\widetilde{\tau}_1} ,\cdots ,\langle \widetilde{A}\rangle_{\widetilde{\tau}_m}\\
         \end{array}
          \right)(\beta_{x,y})
         \left(
          \begin{array}{c}
        \langle \widetilde{B}\rangle_{\widetilde{\omega}_1}\\
         \vdots\\
         \langle \widetilde{B}\rangle_{\widetilde{\omega}_n}\\
        \end{array}
        \right)
       \\ &=\widetilde{\alpha}^T
           \left(
           \begin{array}{c}
           M_{1}\ M_{2}
           \end{array}
           \right)
           \left(
           \begin{array}{c}
           \widetilde{\beta}_1\\ 
           \mathbf{0}
           \end{array}
           \right)\\
           &=\widetilde{\alpha}^T M_{1}\widetilde{\beta}_1\\
           &=0,
\end{align*}
where $\widetilde{\alpha}=(\langle \widetilde{A}\rangle_{\widetilde{\tau}_1}, \cdots, \langle \widetilde{A}\rangle_{\widetilde{\tau}_m})^T$ and $\widetilde{\beta}_1=(\langle\widetilde{B}\rangle_{\widetilde{\omega}_1},\cdots,\langle\widetilde{B}\rangle_{\widetilde{\omega}_l})^T$. Note that we must have $M_1\neq\mathbf{0}$ by the condition ``there exists $i, j $ such that $\beta_{i,j}\Tr(\widetilde{\tau}_i\otimes\widetilde{\omega}_j\cdot \widetilde{A}\otimes\widetilde{B})\neq0$". So similar to the argument before, MDI-EW $((\beta_{x,y}),\{\widetilde{\tau}_x\},\{\widetilde{\omega}_y\},0)$ is sensitive.
    \end{enumerate}
    \end{enumerate}
\end{proof}

\section{Proof of Proposition 5}\label{ap3}
\begin{proof}
By the discussion before, we know that minimizing $\sum\limits_{x,y}\beta_{x,y}\Tr(A_a\otimes B_b\cdot\tau_x\otimes\sigma\otimes\omega_y)$ over all POVM operators $A_a$, $B_b$, all sets of states $\{\tau_x\}$, $\{\omega_y\}$ and all separable states $\sigma$ is equivalent to  
\[
\min\sum\limits_{x,y}\beta_{x,y}\langle A\rangle_{\tau_x}\langle B\rangle_{\omega_y}
\]
over all POVM operators $A$, $B$ and all sets of states $\{\tau_x\}$, $\{\omega_y\}$. Note that when $A$, $B$, $\{\tau_x\}$, $\{\omega_y\}$ are chosen arbitrary, $\langle A\rangle_{\tau_x}$ and $\langle B\rangle_{\omega_y}$ can take any value between 0 and 1 for each $x, y$. So it is equivalent to consider
\[
\min\limits_{\substack{0\leq a_x\leq1 \\ 0\leq b_y\leq1}}\sum\limits_{x,y}\beta_{x,y}a_x b_y.
\]
Denote $f=\sum\limits_{x,y}\beta_{x,y}a_x b_y$, computing the partial derivative $\frac{ \partial f }{ \partial a_x }=\sum\limits_{y}\beta_{x,y}b_y$ which is not depended on $a_x$. Therefore, we know $f$ reaches its minimum on $a_x=0$ or $a_x=1$ for any $x$. Similarly, $f$ reaches its minimum on $b_y=0$ or $b_y=1$ for any $y.$ So we get
\[
\min\sum\limits_{x,y}\beta_{x,y}\Tr(A_a\otimes B_b\cdot\tau_x\otimes\sigma\otimes\omega_y)=\min\limits_{\substack{a_x,b_y\in\{0,1\}}}\sum\limits_{x,y}\beta_{x,y}a_x b_y\equiv\chi.
\]
\end{proof}

\section{Proof of Theorem 6}\label{ap4}
\begin{proof}
There are POVM operators $A_a$, $B_b$ such that
\begin{align*}
\sum\limits_{x,y}\beta_{x,y}p_{\sigma}(a,b\lvert x,y)&=\sum\limits_{x,y}\beta_{x,y}\Tr(A_a\otimes B_b\cdot\tau_x\otimes\sigma\otimes\omega_y)\\
&=\sum\limits_{x,y}\beta_{x,y}\Tr(A_a\otimes B_b\cdot((1-p_A)\widetilde{\tau}_x+p_A\rho_x^{A})\otimes\sigma\otimes((1-p_B)\widetilde{\omega}_y+p_B\rho_y^{B}))\\
&=(1-p_A)(1-p_B)\sum\limits_{x,y}\beta_{x,y}\Tr(A_a\otimes B_b\cdot\widetilde{\tau}_x\otimes\sigma\otimes\widetilde{\omega}_y)+(1-p_A)p_B\sum\limits_{x,y}\beta_{x,y}\Tr(A_a\otimes B_b\cdot\widetilde{\tau}_x\otimes\sigma\otimes\rho_y^{B})\\
&\quad +p_A(1-p_B)\sum\limits_{x,y}\beta_{x,y}\Tr(A_a\otimes B_b\cdot\rho_x^{A}\otimes\sigma\otimes\widetilde{\omega}_y)+p_Ap_B\sum\limits_{x,y}\beta_{x,y}\Tr(A_a\otimes B_b\cdot\rho_x^{A}\otimes\sigma\otimes\rho_y^{B})\\
&\geq(1-p_A)p_B\sum\limits_{x,y}\beta_{x,y}\Tr(A_a\otimes B_b\cdot\widetilde{\tau}_x\otimes\sigma\otimes\rho_y^{B})+p_A(1-p_B)\sum\limits_{x,y}\beta_{x,y}\Tr(A_a\otimes B_b\cdot\rho_x^{A}\otimes\sigma\otimes\widetilde{\omega}_y)\\
&\quad +p_Ap_B\sum\limits_{x,y}\beta_{x,y}\Tr(A_a\otimes B_b\cdot\rho_x^{A}\otimes\sigma\otimes\rho_y^{B})\\
&\geq[(1-p_A)p_B+p_A(1-p_B)+p_Ap_B]\chi\\
\\
&=(p_A+p_B-p_Ap_B)\chi,
\end{align*}
where the first inequality is from $\sum\limits_{x,y}\beta_{x,y}\Tr(A_a\otimes B_b\cdot\widetilde{\tau}_x\otimes\sigma\otimes\widetilde{\omega}_y)\geq0$; and the second inequality is from $\sum\limits_{x,y}\beta_{x,y}\Tr(A_a\otimes B_b\cdot\tau_x^{'}\otimes\sigma\otimes\omega_y^{'})\geq\chi$ for any sets of states $\{\tau_x^{'}\}$, $\{\omega_y^{'}\}$(Proposition \ref{p5}).

Now we deal with the second statement. We know that  inequality \eqref{m3} can detect entanglement in the presence of  noise of the form \eqref{m1}, if and only if
\[
\inf_{\rho}\sum\limits_{x,y}\beta_{x,y}p_{\rho}(a,b\lvert x,y)<(p_A+p_B-p_Ap_B)\chi.
\]
Choose $\tau_x=(1-p_A)\widetilde{\tau}_x+p_A\widetilde{\tau}_x=\widetilde{\tau}_x$, $\omega_y=(1-p_B)\widetilde{\omega}_y+p_B\widetilde{\omega}_y=\widetilde{\omega}_y$($\forall\ x,y$) in \eqref{m1}, and $A_a=\ket{\Phi_{AA}^{+}}\bra{\Phi_{AA}^{+}}$, $B_b=\ket{\Phi_{BB}^{+}}\bra{\Phi_{BB}^{+}}$, where $\ket{\Phi_{AA}^{+}}=\frac{1}{\sqrt{d_A}}\sum\limits_{i=1}^{d_A}\ket{i}\otimes\ket{i}$, $\ket{\Phi_{BB}^{+}}=\frac{1}{\sqrt{d_B}}\sum\limits_{j=1}^{d_B}\ket{j}\otimes\ket{j}$, then
\[
\sum\limits_{x,y}\beta_{x,y}p_{\rho}(a,b\lvert x,y)=\sum\limits_{x,y}\beta_{x,y}\Tr(\ket{\Phi_{AA}^{+}}\bra{\Phi_{AA}^{+}}\otimes\ket{\Phi_{BB}^{+}}\bra{\Phi_{BB}^{+}}\cdot\widetilde{\tau}_x\otimes\rho\otimes\widetilde{\omega}_y)
=\frac{1}{d_Ad_B} \Tr(\sum\limits_{x,y}\beta_{x,y}\widetilde{\tau}_x^T\otimes\widetilde{\omega}_y^T\cdot\rho).
\]
Let $|\Psi\rangle$ be a eigenvector corresponding to the minimal eigenvalue $\lambda_{\min}(\sum\limits_{x,y}\beta_{x,y}\widetilde{\tau}_x^T\otimes\widetilde{\omega}_y^T)$ of $\sum\limits_{x,y}\beta_{x,y}\widetilde{\tau}_x^T\otimes\widetilde{\omega}_y^T$. And set $\rho=|\Psi\rangle \langle \Psi|.$ So we have $\Tr(\sum\limits_{x,y}\beta_{x,y}\widetilde{\tau}_x^T\otimes\widetilde{\omega}_y^T\cdot\rho)=\lambda_{\min}(\sum\limits_{x,y}\beta_{x,y}\widetilde{\tau}_x^T\otimes\widetilde{\omega}_y^T)<0$ which implies that 
\[
\inf_{\rho}\sum\limits_{x,y}\beta_{x,y}p_{\rho}(a,b\lvert x,y)\leq\frac{1}{d_Ad_B}\lambda_{\min}(\sum\limits_{x,y}\beta_{x,y}\widetilde{\tau}_x^T\otimes\widetilde{\omega}_y^T)<0.
\]
  In particular, when $p_A$, $p_B$ satisfy $(p_A+p_B-p_Ap_B)<\frac{1}{d_Ad_B\chi}\lambda_{\min}(\sum\limits_{x,y}\beta_{x,y}\widetilde{\tau}_x^T\otimes\widetilde{\omega}_y^T)$, as $\chi<0$, we have
\[
\inf_{\rho}\sum\limits_{x,y}\beta_{x,y}p_{\rho}(a,b\lvert x,y)\leq\frac{1}{d_Ad_B}\lambda_{\min}(\sum\limits_{x,y}\beta_{x,y}\widetilde{\tau}_x^T\otimes\widetilde{\omega}_y^T)<(p_A+p_B-p_Ap_B)\chi.
\]
\end{proof}


\begin{thebibliography}{99}
\bibitem{1}  R. Horodecki, P. Horodecki, M. Horodecki and K. Horodecki, Quantum entanglement, \href{https://link.aps.org/doi/10.1103/RevModPhys.81.865}{Rev. Mod. Phys. \textbf{81}, 865 (2009)}.
\bibitem{2} O. G{\"u}hne and G. T{\'o}th, Entanglement detection, \href{https://doi.org/10.1016/j.physrep.2009.02.004}{Phys. Rep. \textbf{474}, 1 (2009)}.
\bibitem{3} N. Friis, G. Vitagliano, M. Malik, and M. Huber, Entanglement certification from theory to experiment, \href{https://doi.org/10.1038/s42254-018-0003-5}{Nat. Rev. Phys. \textbf{1}, 72-87(2019)}
\bibitem{4} B.M. Terhal, Bell inequalities and the separability criterion, \href{https://doi.org/10.1016/S0375-9601(00)00401-1}{Phys. Lett. A \textbf{271}, 319 (2000)}.
\bibitem{5} M. Horodecki, P. Horodecki, and R. Horodecki, Separability of mixed states: necessary and sufficient conditions, \href{https://doi.org/10.1016/S0375-9601(96)00706-2}{Phys. Lett. A \textbf{223}, 1 (1996)}.
\bibitem{6} D. Chru{\'s}ci{\'n}ski, G. Sarbicki, Entanglement witnesses: construction, analysis and classification, \href{https://doi.org/10.1088/1751-8113/47/48/483001}{J. Phys. A: Math. Theor. \textbf{47}, 483001 (2014)}.
\bibitem{7} M. Seevinck and J. Uffink, Local commutativity versus bell inequality violation for entangled states and versus non-violation for separable states, \href{https://doi.org/10.1103/PhysRevA.76.042105}{Phys. Rev. A \textbf{76}, 042105 (2007)}.
\bibitem{8} D. Rosset, R. Ferretti-Schöbitz, J.-D. Bancal, N. Gisin, and Y.-C. Liang, Imperfect measurement settings: Implications for quantum state tomography and entanglement witnesses, \href{https://doi.org/10.1103/PhysRevA.86.062325}{Phys. Rev. A \textbf{86}, 062325 (2012)}.
\bibitem{9} S. Morelli, H. Yamasaki, M. Huber, and A. Tavakoli, Entanglement detection with imprecise measurements, \href{ https://doi.org/10.1103/PhysRevLett.128.250501}{Phys. Rev. Lett \textbf{128}, 250501 (2022)}.
\bibitem{10} S. Sarkar, Distrustful quantum steering \href{https://doi.org/10.1103/PhysRevA.108.L040401}{Phys. Rev. A \textbf{108}, L040401 (2023)}.
\bibitem{11} A. Tavakoli, Quantum steering with imprecise measurements, \href{https://doi.org/10.1103/PhysRevLett.132.070204}{Phys. Rev. Lett \textbf{132}, 070204 (2024)}.
\bibitem{12} H. Cao, S. Morelli, L.A. Rozema, C. Zhang, A. Tavakoli, and P. Walther, Genuine Multipartite Entanglement Detection with Imperfect Measurements: Concept and Experiment, \href{https://doi.org/10.1103/PhysRevLett.133.150201}{Phys. Rev. Lett \textbf{133}, 150201 (2024)}.
\bibitem{13} C. Branciard, D. Rosset, Y.C. Liang, N. Gisin, Measurement-device-independent entanglement witnesses for all entangled quantum states, \href{https://doi.org/10.1103/PhysRevLett.110.060405}{Phys. Rev. Lett \textbf{110}, 060405 (2013)}.
\bibitem{14} K. Sen, C. Srivastava, S. Mal, A. Sen, U. Sen, Noisy quantum input loophole in measurement-device-independent entanglement witnesses, \href{https://doi.org/10.1103/PhysRevA.104.012429}{Phys. Rev. A \textbf{104}, 012429 (2021)}.
\bibitem{15} M. Lewenstein, B. Kraus, J.I. Cirac, and P. Horodecki, Optimization of entanglement witnesses, \href{https://doi.org/10.1103/PhysRevA.62.052310}{Phys. Rev. A \textbf{62}, 052310 (2000)}.
\bibitem{16} F. Buscemi, All entangled quantum states are nonlocal, \href{https://doi.org/10.1103/PhysRevLett.108.200401}{Phys. Rev. Lett \textbf{108}, 200401 (2012)}.
\bibitem{17} E. Svegborn, N. d'Alessandro, O. G{\"u}hne, and A. Tavakoli, Imprecision plateaus in quantum steering, \href{https://doi.org/10.48550/arXiv.2408.12280}{arXiv: 2408.12280 (2024)}.
\bibitem{18} R.F. Werner, Quantum states with Einstein-Podolsky-Rosen correlations admitting a hidden-variable model, \href{https://doi.org/10.1103/PhysRevA.40.4277}{Phys. Rev. A \textbf{40}, 4277 (1989)}.
\bibitem{19} A. Peres, Separability Criterion for Density Matrices, \href{https://doi.org/10.1103/PhysRevLett.77.1413}{Phys. Rev. Lett. \textbf{77}, 1413 (1996)}.
\bibitem{20} G. T{\'o}th and O. G{\"u}hne, Detecting Genuine Multipartite Entanglement with Two Local Measurements, \href{https://doi.org/10.1103/PhysRevLett.94.060501}{Phys. Rev. Lett. \textbf{94}, 060501 (2005)}.
\bibitem{21} N. Brunner, D. Cavalcanti, S. Pironio, V. Scarani, and S. Wehner, Bell nonlocality, \href{https://doi.org/10.1103/RevModPhys.86.419}{Rev. Mod. Phys. \textbf{86}, 419 (2014)}.
\bibitem{22} J.-D. Bancal, N. Gisin, Y.-C. Liang, and S. Pironio, Device-independent witnesses of genuine multipartite entanglement, \href{https://doi.org/10.1103/PhysRevLett.106.250404}{Phys. Rev. Lett. \textbf{106}, 250404 (2011)}.
\bibitem{23} J. T. Barreiro, J.-D. Bancal, P. Schindler, D. Nigg, M. Hennrich, T. Monz, N. Gisin, and R. Blatt, Demonstration of genuine multipartite entanglement with device-independent witnesses, \href{https://doi.org/10.1038/nphys2705}{Nature Phys \textbf{9}, 559 (2013)}.

\end{thebibliography}
\end{document}